\documentclass[twocolumn,showpacs,amsmath,amssymb,prl]{revtex4}
\usepackage{graphicx}% Include figure files
\usepackage{dcolumn}% Align table columns on decimal point
\usepackage{bm}% bold math
\usepackage{aas_macros}
\usepackage[colorlinks=true,linkcolor=blue,citecolor=blue]{hyperref} % Matches PRL style
\begin{document}
\input{epsf}

\title{Observational Cosmology With Semi-Relativistic Stars}

\author{Abraham Loeb and James Guillochon}

\affiliation{Institute for Theory \& Computation, Harvard University,
60 Garden St., Cambridge, MA 02138, USA}

\begin{abstract}

Galaxy mergers lead to the formation of massive black hole binaries
which can accelerate background stars close to the speed of light. We
estimate the comoving density of ejected stars with a peculiar
velocity in excess of $0.1c$ or $0.5c$ to be $\sim 10^{10}$ and $10^5$
$~{\rm Gpc^{-3}}$ respectively, in the present-day
Universe. Semi-relativistic giant stars will be detectable with
forthcoming telescopes out to a distance of a few Mpc, where their
proper motion, radial velocity, and age, can be spectroscopically
measured.  In difference from traditional cosmological messengers,
such as photons, neutrinos, or cosmic-rays, these stars shine and so
their trajectories need not be directed at the observer for them to be
detected.  Tracing the stars to their parent galaxies as a function of
speed and age will provide a novel test of the equivalence principle
and the standard cosmological parameters. Semi-relativistic stars
could also flag black hole binaries as gravitational wave sources for
the future {\it eLISA} observatory.

\end{abstract}

\date{\today}
\maketitle

\paragraph*{Introduction.}
Mergers of galaxies produce binary systems of massive black holes
embedded in a dense environment of stars \cite{Colpi}. Such binaries
are capable of accelerating rare background stars to high speeds
\cite{Hills,Yu,Levin,OL}. Here we identify the conditions under which
the accelerated stars reach a fraction of the speed of light, $\gtrsim
0.1c$, and calculate the cosmological implications of the resulting
population of semi-relativistic hypervelocity stars (SHS) that fills
the Universe.
 
\paragraph*{SHS abundance.}

N-body simulations suggest that massive black hole binaries are
excited to high eccentricities prior to merger
\cite{Sesana,Iwasawa,Khan}. Stars originally bound to the less massive
black hole (secondary) can then be ejected at high speeds, in a manner
similar to the production of hypervelocity stars from stellar binaries
orbiting a single black hole \cite{Hills}. The ejection speed can
exceed the orbital speed of the stars \cite{Sari}, occasionally
ejecting SHS near the speed of light. The highest speeds are achieved
when stars approach the secondary black hole as closely as possible
(without being tidally disrupted) at the moment of its closest
approach to the primary black hole. The likelihood of ejecting stars
at different speeds depends on the galaxy merger rate, the mass ratio
of the resulting binaries, and the distribution of stars around the two
black holes during the mergers.

\begin{figure}
\centering\includegraphics[width=\linewidth,clip=true]{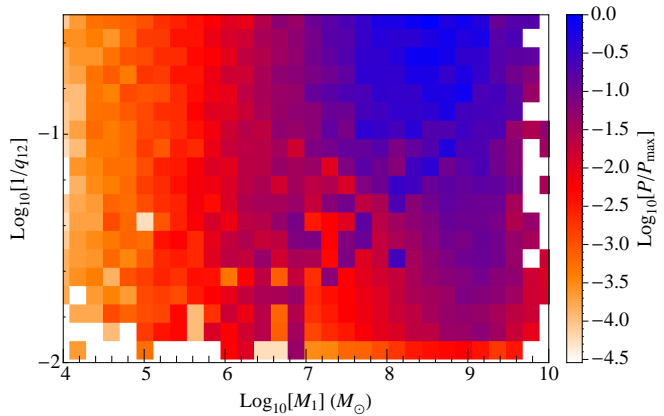}
\caption{Fraction of SHS $P/P_{\max}$ that originate from binary black hole
  binaries of a given primary mass $M_{1}$ and primary-to-secondary
  ratio $q_{12}$, where $P_{\max}$ is the maximum probability amongst all mass combinations}
\label{fig:masses}
\end{figure}

We have conducted numerical calculations of the statistics of SHS,
whose details are reported in a companion paper \cite{GL}.  Figure~\ref{fig:masses}
shows the fraction of SHS originating from black hole binaries
of a given primary mass $M_1$ and primary-to-secondary mass ratio
$q_{12}$, and illustrates the fact that most of these stars originate
from the mergers of the very largest black holes in the universe
($M_{1} \sim 10^{8}$--$10^{9} M_{\odot}$).

\begin{figure}
\centering\includegraphics[width=0.9\linewidth,clip=true]{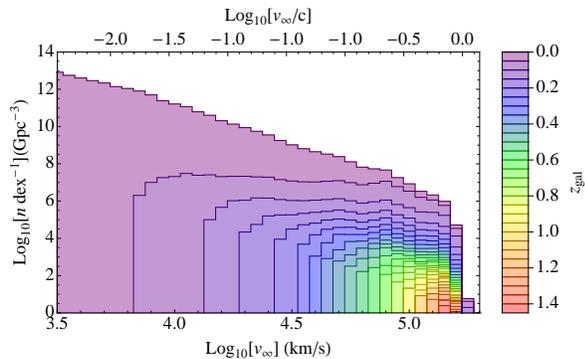}
\caption{Number density of SHS $n$ originally ejected at velocity $v_{\infty}$, normalized per decade of $\log_{10} v_{\infty}$. The colored shading indicates the present-day redshift of the source galaxy $z_{\rm gal}$, with higher velocity SHS typically originating from more distant galaxies.}
\label{fig:hist}
\end{figure}

The resulting comoving density of SHS with different speeds in the
present-day universe \footnote{The statistics of stars very close to
  the speed of light requires an extension of our simplified Newtonian
  calculation to General Relativity, and should depend on the unknown
  distribution of black hole spins.} is illustrated in Fig.~\ref{fig:hist},
and shows that tens of trillions of such stars lie within each cubic Gpc, many
of which originate from galaxies that are at redshifts exceeding unity.  In
what follows, we focus on the implications of the existence of these
stars for cosmological studies.

\begin{figure}
\centering\includegraphics[width=0.85\linewidth,clip=true]{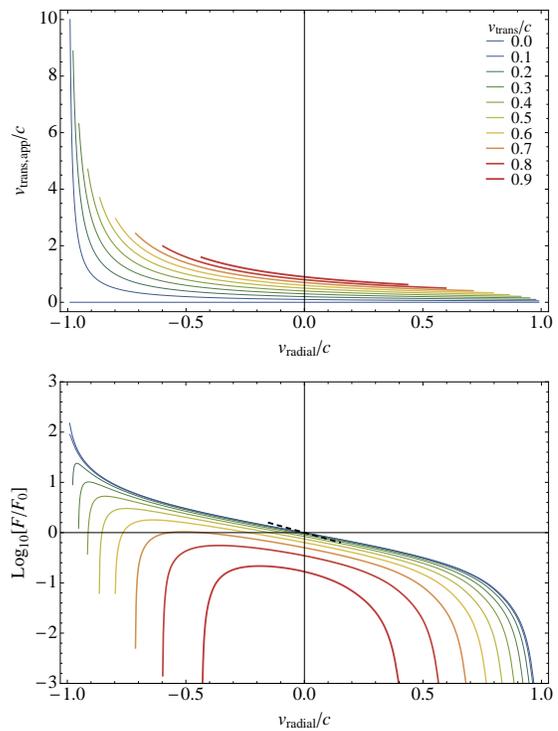}
\caption{{\it Upper panel:} The apparent transverse speed of an SHS as
  a function of its actual transverse velocity $v_{\rm trans}$ and
  radial velocity $v_{\rm radial}$ (in units of $c$). {\it Lower
    panel:} Boost in total flux $F$ compared to its value at zero
  velocity $F_0$ for an M3III giant SHS as a function of the radial
  ($v_{\rm radial}$) and transverse ($v_{\rm trans}$) components of
  its velocity. The dashed line segment shows the expected change in
  flux for a flat spectrum and purely radial motion when $v\ll c$.}
\label{fig:relativity}
\end{figure}

The intrinsic SHS flux per unit frequency $F_\nu$ would be observed to
be boosted by a factor $D^{3-\alpha}$, where $D\equiv
[\gamma(1-\beta\cos\theta)]^{-1}$ is the relativistic Doppler factor
for a SHS propagating with a speed $\beta=(v/c)=\sqrt{1-\gamma^{-2}}$
at a angle $\theta$ relative to the line-of-sight, and $\alpha=d\ln
F_\nu/d\ln \nu$ is the average spectral index in the observed
frequency band \cite{RL}. For a blackbody stellar spectrum with an
effective temperature $T_{\rm eff}$,
$\alpha(\nu)=[{e^{x}(3-x)-3}]/[{e^{x}-1}]$, where $x\equiv
{h\nu/kT_{\rm eff}}$.  In the Rayleigh-Jeans part of the spectrum ($x
\ll 1$), $\alpha=2$, whereas in the Wien tail ($x\gg 1$),
$\alpha=3-x$.  The apparent transverse speed \cite{Rees} of a SHS,
$\beta_{\perp,app}=(D\sin\theta) \gamma\beta$, peaks at a value of
$\gamma\beta$ (which could exceed unity) for $\cos
\theta=\beta$. Figure~\ref{fig:relativity} shows the expected boost in the observed flux
and proper motion of an M3III giant star as functions of the radial
and tranverse components of its velocity. Different stellar spectra
lead to different average values of $\alpha$ (depending on the
observed frequency band) and hence different boost factors.

\paragraph*{Trajectory in a homogeneous Universe.}
Next, we consider the trajectory of a SHS that reached the vicinity of
the Milky-Way galaxy at the present time $t_0$ with a measured speed
$v_0$. The peculiar momentum per unit mass of any object in excess of
the Hubble flow $\gamma v$ declines inversely with the scale-factor
$a(t)=(1+z)^{-1}$ in an expanding universe (since the de Broglie
wavelength is stretched like any other physical length scale).  In the
non-relativistic regime, this implies that the peculiar velocity of
the star at earlier times $t<t_0$ was,
\begin{equation}
v=v_0/a(t).
\label{eq:1}
\end{equation}
On large spatial scales the Universe is filled with a nearly uniform
distribution of SHS, with the velocity dispersion declining as
$a^{-2}$, as expected for a collisionless gas of non-relativistic
particles.

To leading order, we assume a homogeneous, isotropic and 
geometrically
flat Universe described by the metric,
$ds^2=c^2dt^2-a^2(t)(dr^2+r^2d\Omega)$, where $r$ is the comoving
(present-day) radius relative to the observer. For simplicity, we
consider the regime where the source galaxy from where the star was
ejected at time $t_{ej}$ is much farther away ($r\sim {\rm Gpc}$) than
the distance of the star from the observer ($\sim {\rm Mpc}$) which is
itself much larger than the eventual distance of closest approach
relative to the observer. In this regime, the trajectory of the star
is nearly radial towards the observer, with a time-dependent velocity
$v(t)=a(t)dr/dt$. By substituting this relation in Eq. (\ref{eq:1}),
we get $dr=v_0 dt/a^2(t)$, and after integrating both sides of this
equation we find the comoving distance of the source galaxy,
\begin{equation}
r=v_0 \int_{t_{ej}}^{t_0} {dt \over a^2(t)} .
\label{eq:2}
\end{equation}
Note that SHS with different $v_0$ offer the opportunity to measure
the distances of a source galaxy at multiple look-back times, in difference
from photons which provide only a single snapshot due to their fixed propagation speed.
The value of $r$ as a function of the star's travel time
$t_\star=(t_0-t_{ej})$ is shown by the upper curve in Fig.~4 for the
standard LCDM cosmology. The SHS-producing merger of galaxies often funnels gas to the galaxy centers and results in star formation there. Many of the newly formed stars are tightly bound to their black holes, and could have been slingshot ejected when they were much younger than their age today.  In general, the measured age of each SHS only represents the maximum travel time that it could have had. But among all the stars that move at the same speed and have the same age, those that traverse the largest distance are most likely to have been ejected at youth. As long as multiple such stars provide consistent cosmological constraints, their use as rulers would be reliable.

\begin{figure}[th]
\includegraphics[scale=0.3]{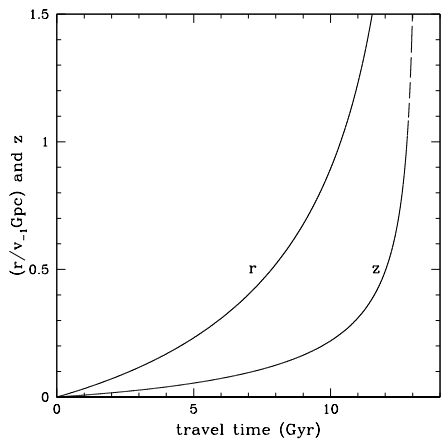}
\caption{The upper curve shows the comoving distance $r$ of the 
source
  galaxy in units of $v_{-1}$Gpc [where $v_{-1}\equiv (v_0/0.1c)$] as
  a function of the star's travel time $t_\star$ in Gyr for the
  standard LCDM cosmology \cite{Planck} with $\Omega_m=0.32$,
  $\Omega_\Lambda=0.68$ and $H_0=67~{\rm km~s^{-1}~Mpc^{-1}}$.  
The
  lower curve shows the corresponding source redshift in the
  particular case of $v_0=0.1c$; lower $v_0$ would translate to lower
  redshifts. At short distances, $z\propto r\propto v_0$.}
\label{fig:2}
\end{figure}

\begin{figure}
\centering\includegraphics[width=0.9\linewidth,clip=true]{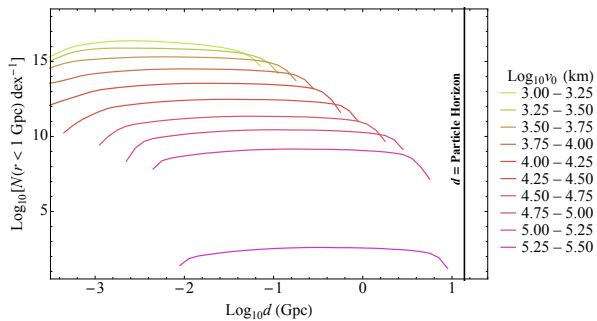}
\caption{Histograms of distances traveled $d$ by SHS observed within 1
  Gpc as a function of their present-day speed $v_{0}$.}
\label{fig:dist}
\end{figure}

In difference from traditional cosmological observations in which
$v\approx c$ messengers (such as photons, neutrinos, or cosmic rays)
must intersect the collecting area of a telescope in order to be
detected, SHS can be detected at a distance. While photons must follow
radial trajectories from the source to the observer, SHS emit their
own light and therefore can be detected even if they miss the observer
with a finite impact parameter or pass beyond the observer's
location. This implies that SHS could possess substantial proper
motion near their point of closest approach.  On average, one expects
equal numbers of redshifted and blueshifted SHS that move at a fraction of $c$; the former class
involving stars that have already passed through (or were just ejected
out of) the observer's region. Figure~\ref{fig:dist} shows the distances traveled
by an SHS observed within 1 Gpc, as a function of its observed speed.

\paragraph*{Gravitational deflection.} 
A non-relativistic star passing within an impact parameter $b$ from a
projected (cylindrically-integrated) concentration of mass $M(b)$
along its path, would be deflected by an angle \cite{BT},
\begin{equation}
\theta={2GM(b)\over v^2 b}={2V_c^2\over
v^2}=27.5^{\prime\prime}\left[{(V_{c}/200~{\rm km~s^{-1}})\over
(v/0.1c)}\right]^2 ,
\label{eq:3}
\end{equation}
where $V_c\equiv \sqrt{GM(b)/b}$ \footnote{Note that the deflection
  angle of a non-relativistic particle deviates from that of a photon
  by a factor of 2 aside from the substitution of $v$ for $c$.}. The
deflection would also lead to a slight delay in arrival time relative
to the relation between $r$ and $t_\star$ in Fig.~\ref{fig:2}.  If the
parent galaxy is identified, measuring the deflection angle of SHS as
a function of speed can be used to determine the projected mass profile
of the deflector, be it a galaxy or a cluster of galaxies.

The angular cross-section $A$ for deflecting a star ejected from a
source galaxy behind a singular isothermal sphere with a 1D velocity
dispersion $\sigma$, can be derived in analogy with gravitational
lensing of photons \cite{LF},
\begin{equation}
A=4\pi^3 \left({\sigma\over v}\right)^4 \left(1-{r_d\over
r_s}\right)^2 ,
\label{lens}
\end{equation}
where the subscripts $d$ and $s$ label the deflector and source,
respectively. This cross-section is larger by a factor of ${1\over 4}
(c/v)^4$ than the Einstein cross-section for gravitationally-lensed
photons \cite{Keeton}.  At low $v$, the maximum value of $A$ is
limited by the finite virial radius of the deflector, $R_{vir}$. In
the following, we consider a sufficiently large $v$ such that
$b<R_{vir}$. For short travel times $t_\star\ll H_0^{-1}=14~{\rm
Gyr}$, the optical depth for deflection $\tau$ is obtained by
integrating the local density of deflectors times $A$ over the path
length, giving
\begin{equation}
\tau=4\pi^3 \Gamma n_\star \left({\sigma_\star \over
v_0}\right)^4r_s^3\approx 0.2 \left({v_0\over
0.1c}\right)^{-4}\left({r_s\over {\rm Gpc}}\right)^3 ,
\end{equation}
where for the local galaxy population we have adopted the
Faber-Jackson relation $L/L_\star=(\sigma/\sigma_\star)^\gamma$ and 
a
Schechter luminosity function to describe the differential number
density of galaxies per luminosity $L$ bin,
$dn/dL=(n_\star/L_\star)(L/L_\star)^\alpha\exp\{-L/L_\star\}$, with
the parameters $n_\star= 5\times 10^{-3}~{\rm Mpc^{-3}}$,
$\sigma_\star= 130~{\rm km~s^{-1}}$, $\alpha=-1.05$, $\gamma= 4.1$
\cite{Cool,Brim}, and $\Gamma\equiv\Gamma(1+\alpha+{4/\gamma})$.  
The
likelihood for SHS deflection by galaxies and large-scale structure
\cite{Ref} exceeds unity at low velocities, $v_0\lesssim 0.07
(r_s/{\rm Gpc})^{3/4} c$. For long travel times, it is necessary to
incorporate Eq. (\ref{eq:1}) into the deflection cross-section in
calculating the optical depth.

If the star was ejected when it was much younger than its present age,
then one could infer $t_{ej}=t_0-t_\star$ from a spectroscopic
measurement of the star's age. Forthcoming surveys, such as GAIA \cite{GAIA} or LSST \cite{LSST}, 
and future observatories, such as JWST \cite{JWST} or large-aperture
ground-based telescopes \cite{ELT}, could detect SHS during the giant
phase of their evolution out to distances of order a few Mpc
\cite{GL}. Follow-up spectroscopy could measure the radial velocity
and age of SHS, which when combined with proper motion, can be
used to relate SHS to their parent galaxies.

\paragraph*{Identifying the parent galaxy.}
If deflections along the star's trajectory can be corrected for, and
the source galaxy can be identified from a spectroscopic determination
of the age of the star ($\gtrsim t_\star$), its
spectroscopically-measured radial velocity ($v_0$) and its observed
proper motion, then one might be able to uniquely identify the source
galaxy and measure its redshift, $z$.  The cosmological evolution of
the scale factor, $a(t)=(1+z)^{-1}$, depends on cosmological
parameters through the Friedmann equation. In principle, if the
measurement accuracy is sufficiently high, it may be possible to
constrain cosmological parameters (such as the equation of state of
dark energy) by requiring that the source redshift (as measured by
detecting photons at $t_0$) would agree with the comoving distance
traveled by a star with a present-day velocity $v_0$ over its travel
time $t_\star$ (cf. Eq. \ref{eq:2}). The photon redshift
$z=a^{-1}(t_\gamma)-1$ is obtained from the equation,
\begin{equation}
r=c\int_{t_\gamma}^{t_0} {dt \over a(t)} =\int_{1/(1+z)}^1 {da \over a^2H},
\label{eq:4}
\end{equation}
where $H=H_0\sqrt{\Omega_m a^{-3}+\Omega_\Lambda}$ is the 
Hubble
parameter for the redshifts of interest, with $\Omega_m$ and
$\Omega_\Lambda$ being the present-day cosmological parameters of 
the
matter and vacuum.  By comparing Eqs. (\ref{eq:2}) and (\ref{eq:4}),
it is clear that the photons must have been emitted after the SHS left
the galaxy, in order for them to arrive to the observer at the same
time. The lower curve in Fig.~\ref{fig:2} shows the source redshift as
a function the star's travel time $t_\star$ for the particular case of
$v_0=0.1c$. This curve (as a function of $v_0$) provides a new variant
of the conventional ``Hubble Diagram'' which traditionally uses
photons as messengers. Comparing SHS with photons from the same 
source
galaxy would constitute a novel test of the equivalence principle and
the standard model of cosmology.

\paragraph*{Intergalactic light.}
In addition to individual SHS passing relatively close to the observer,
the combined effect of all SHS in the observable universe can contribute
a significant fraction of the light in the voids between galaxy clusters. Most void light originates
from within a few hundred kpc of the galaxies occupying those voids \cite{Deason},
and thus away from these void galaxies SHS are likely to be a signifcant source of light.
We calculate that the intergalactic interhalo light from all SHS in the Universe (which constitute
$\sim 10^{-5}$ of all stars at $z=0$) is $\sim 10^{-3}~{\rm nW~m^{-2}~sr^{-1}}$ at a wavelength
of 1$\mu$m. This is comparable to the minimum value expected in some regions of the sky \citep{Croton}.

\paragraph*{Broader implications.}
The existence of SHS has multi-disciplinary implications. In the
context of astro-biology, SHS could spread life beyond the boundaries
of their host galaxies \cite{Pan,Loeb}. In the context of
gravitational-wave astrophysics, the abundance of SHS can be used to
calibrate the coalescence rate of tight black holes binaries, especially the more-massive examples which produce the majority of SHS (see Figure \ref{fig:masses}), and for which some examples ($M_{1} \sim 10^{8} M_{\odot}$) have been discovered recently \citep{Valtonen,Graham}.  If the parent galaxies are identified, SHS could flag additional binary black holes systems that would be strong
gravitational wave sources for the future {\it e-LISA} observatory
\cite{LISA}.

\bigskip
\bigskip
\paragraph*{Acknowledgments.}
This work was supported in part by NSF grant AST-1312034 (A.L.) and Einstein grant PF3-140108 (J.G.).

\end{document}